# Data Gathering from Path Constrained Mobile Sensors Using Data MULE

Dinesh Dash, NIT Patna, India
dd@nitp.ac.in

**Abstract**—In Wireless Sensor Network (WSN) sensor nodes are deployed to sense useful data from environment. Sensors are energy-constrained devices. To prolong the sensor network lifetime, now a days mobile robots (sometimes refer as data sink, data mules, or data collectors) are used for collecting the sensed data from the sensors. In this environment sensor nodes directly transfer their sensed data to the data mules. Sensed data are sometime time sensitive; therefore, the data should be collected within a predefined period. Hence, depending on the speed of the data mules the trajectory lengths of the data mules have upper limits. In this paper an approximation algorithm is proposed for collecting data from the mobile sensors using mobile data collectors.

## 1. INTRODUCTION

Wireless Sensor Network (WSN) consists of large number of sensors (nodes) and few base stations (BS). Each sensor has a sensing range and a communication range. Within the sensing range sensor can sense environmental data and it can communicate to other sensors which are within its communication range. A typical application in WSN is to collect the sensed data from individual sensors to a BS. Depending on the communication range of the sensors, they form a communication network topology. Two nodes are connected by an edge if they are within their communication range. Most of the nodes cannot communicate directly to the BS and they send data packet to BS through multi-hop communication.

Hierarchical or cluster-based routing methods are proposed in wireless networks, in which a subset of node is selected to form communication layer topology, and only the nodes in the communication layer participate for data communication and thereby reduce the transmission overhead of redundant information. It simplifies the topology of the network and saves energy for information gathering and forwarding.
Data collection is one of the fundamental operations in WSN. Other critical network operations such as event detection, robust message delivery, localization, network reconfiguration etc. are depended on data collection as a basic operation. Data aggregation and in-network processing techniques have been investigated recently as efficient approaches to achieve significant energy savings in WSN by combining data arriving from different sensor nodes at some aggregation points, eliminating redundancy, and minimizing the number of transmission before forwarding data to the sinks. Hence, data fusion or aggregation has emerged as a useful paradigm in sensor networks. Due to the multi-hop data transmission in static sink based WSN, unbalanced energy consumption is caused to the nodes close to sink and other distant sensor nodes. Sensor nodes, which are close to the sink node, have to carry much more traffic overhead compared with distant sensor nodes. Since sensor nodes are limited with battery power supply, such unbalanced energy consumption causes quick power depletion on part of the network, and reduced the lifetime of the network. To resolve this issue recent research works propose mobile sink based data gathering techniques.

Mobile sink is an unmanned vehicle/ robot that roam around the area and collects sensed data from data collectors. Mobile sink based data gathering techniques are also useful in applications involving real-time data traffic. In such applications, data gathering paths by the mobile sink are selected so that certain end-to-end delay constraints are satisfied. In order to improve the round-trip time of the mobile sink, one of the solutions could be to move the sink only to few data collectors rather than individual sensors. To fulfil the time constraint of real time sensor data a subset of sensors called s cluster heads are selected efficiently so that the total length to visit them for collecting the data is minimum.

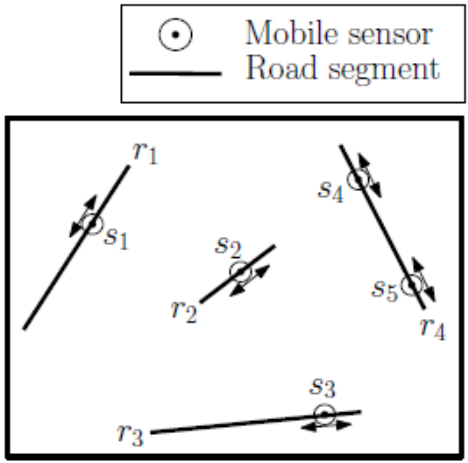

Figure 1

In path constraint mobile sensor network, sensors are moving along a set of pre-defined paths (roads). Mobile sink can move to any arbitrary position for collecting data. An example of path constrained mobile sensor network is shown in Figure 1. Paths are denoted by $\{r_1, r_2, .. r_4\}$ (paths for the mobile sensors), circles $\{s_1, s_2, … s_5\}$ denote mobile sensors. The mobility of the sensors are confined with the segments.

Data gathering problem from mobile sensors is a challenging problem

### 1.1 Contribution
In this paper, our contributions on mobile data gathering problem are as follows:

- We have addressed periodic data gathering protocol from a set of mobile sensors where trajectory of the mobile sensors are confined within a set of straight line segments on a plane.
- We have identified two limitations of the solution proposed in [9] and proposed a solution to overcome the limitations.
- A 4 approximation algorithm is proposed for the above problem.

The rest of the paper is organized as follows. Section 2 briefly describes related work of the paper. Formal definition of problem and network model is described in Section 3. Our solution approach is discussed in Section 4. Section 5 concludes the paper and describes some of its future works.

## 2. RELATED WORKS

In wireless sensor networks data gathering protocols are broadly classified into two types static sink node based and mobile sink node based. In static sink architecture all static sensor send the sensed data to the static sink node through multi-hop communication where as in mobile sink architecture the sensed data are collected by a mobile sink node by visiting the communication range of individual sensors after a certain time interval.

In static sink architecture, there are two types of nodes: regular node and sink node. Regular node senses the data and sends the data to the sink node. There is a data aggregation tree rooted at the sink. The tree is generated from the communication topology of the sensor network. He et al. [7] proposed a load balanced data aggregation tree for gathering the sensed from a probabilistic network model.

In mobile sink based data gathering protocols, the mobile sink instead of visiting all static sensors, chooses a sub set of sensors as gateways. Therefore, in this type network, there are three types of sensor nodes: regular node, intermediate gateway node and sink node. Regular nodes sense the environmental data and send it either to a sink or to an intermediate gateway node. Gateway node works as regular node as well as it also helps to forward other sensors data.

In [2], Liu et al. proposed a cluster based solution for finding optimal path of mobile sink to collect the data from the individual sensors. In this protocol there is no restriction on path of the mobile sink. The sensors with common intersection on their communication range forms a cluster. The mobile sink only visit the common intersection zones to collect the sensed data. It will collect the sensed data directly from each individual sensor so there is no intermediate gateway node. The total trajectory of the mule is minimum. An improve version of the genetic algorithm is proposed in [4]. In this works authors proposed a novel method for population generation in the genetic algorithm and there after effective shortcuts techniques are proposed to improve the path length.

In [3], Kim et al. proposed approximation algorithm to collect data from static sensors using data mules. Data mules collect data from the sensor within its neighbourhood. To reduce the data gathering paths of the data mules multiple data mules are used. To reduce data gathering period maximum trajectory length of the data mules is minimized. Gao et al. in [6], present a data-gathering algorithm from a set of static sensor nodes using a mobile data collector. Path of the mobile data collector (sink) is a fixed path P. Objective of the work is to find a continuous sub path P' on the path P of length V*T/2 for mobile sink node, where V is velocity of the mobile sink and T is the time deadline of the sensed data. Such that the total energy consumption by the network to deliver the data to the sub sink nodes closed to the path P' is minimum. Sub-path based on the maximum number of sub sink closed to the path within predefined travel distance. In [1], Mai et al. proposed a load balance mobile sink based data gathering protocol where the mobile sink runs with a given speed and wants to finish its data gathering cycle within a given time bound. They assumed that there is no restriction on the path of the mobile sink. The objective is to select a set of gateways such that loads on the gateways are almost same and the trajectory length of the mobile sink satisfies the maximum delay.

Gao et al. [5] proposed a path constrained data gathering protocol. They try to maximize the amount of data collected per round of the mobile sink. As the mobile sink roam around a predefined path with a fixed speed, it will stay for a fixed amount of time close to any intermediate gateway node. Hence if the intermediate gateway node has too many data it will be unable to transfer the whole data to the mobile sink within that time period. In order to increase the total amount of gathered data by the mobile sink, the data sensed by sensors must be uniformly distributed amongst the gateways. ILP formulation of the problem is shown and a genetic algorithm is proposed. In [8], a subset of sensors is considered as a data source all of them generate same amount data. Different sensors have different time deadline, within the time deadline data need to be delivered to a predefined sink node. Goal is to find energy efficient set of paths from the sources to the sink, which will satisfy the time deadline as well as the total energy consumption is minimum.

Data gathering from mobile sensors is another challenging area in sensor network. In [9], a solution for data gathering from path constrained mobile sensors is proposed using data mules. The paths of the mobile sensors are restricted along a set of line segments and the mobile sink visits all the mobile sensors to collect their sensed data.

### 3. PROBLEM STATEMENT AND COMPLEXITY

A set of mobile sensors S= {$s_1$, $s_2$,… $s_N$} are moving along a set of road segments R= {$r_1$, $r_2$, … $r_M$}. Assume a data mules can collect data from a mobile sensor when it reaches to the point where mobile sensor presents. Assume movement paths of the mobile sensors are arbitrary along the R. Movement speed of the mobile sensors are also arbitrary. Sensors can stop movement for arbitrary time. Data mules can move to any location on plane and moving with a fixed speed V.

**Problem 1:** Find minimum number of data mules and their travelling paths to collect the sensed data from all mobile sensors within a specified time period t.

### 4. ALGORITHM TO FIND MINIMUM DATA MULES AND THEIR PATHS

In this section, we discuss a data gathering protocol using mobile data collectors to collect data from a set of mobile sensors which are moving on a set of road segments. We refer the algorithm as data gathering from path constrained mobile sensors (DGPCMS). Gorain et al. [9] propose an approximation algorithm for data gathering from path constraint mobile sensors. But the proposed solution has few limitations such as (i) the length of the road segments are bounded by Vt/2, where V is the speed of the data mules and t is the data gathering time interval, and (ii) the data from the mobile sensors are collected by all the data mules. In this paper, we have addressed these two limitations. And propose an approximation solution for the problem, which is based on the following characteristic.

**Lemma 1:** Every point of all road segments must be visited by a data mule at least once within time interval t.
**Proof:** A mobile sensor can move to any arbitrary position on its predefined road segment. Therefore, at least one data mule must visit every point of all the line segments within time interval t.

Determine the shortest distance $c_{ij}$ between end points of every pair of segments ($r_i$, $r_j$) where i≠j and call them as inter segments connectors. Construct a complete graph among the segments with the help of inter segments connectors. Our algorithm determines a solution of the number of data mules requirement in M rounds, where M is the number road segments. In particular round k (k varying in between 1 to M) finds a minimum spanning forest $F_k$ with (M-k+1) trees by interconnecting the segments in R with (k-1) inter segments connectors. For each tree $T_i \in F_k$, i ∈{1 to k} construct Euler cycle $EC_i$ by doubling each and remove the highest length inter segments connector from $EC_i$ to convert it to Euler path $E_i$. Let $N_k$ denote an upper bound on the number of data mules requirement by our algorithm to traverse (M-k+1) Euler paths in $k^{th}$ round. Divide (M-k+1) Euler paths into segments of equal length of size at most V$t$. Thereafter deploy two data mules at the two ends of every segment and they start moving towards each other until they meet at the middle and thereafter reverse their movement direction until they reach their starting points again. Since the length of Euler path $L(E_i) \leq 2L(T_i)$ for i∈{1, 2,… (M-k+1)}, we can write $N_k \leq 2 \sum_{i=1}^{M-k+1} \left\lceil \frac{2L(T_i)}{Vt} \right\rceil$, where $L(T_i)$ denotes length of tree $T_i$ and V$t$ denotes the distance travel by a data mule within t time period. The detail algorithm is shown in Algorithm 1.

---

**Algorithm 1:** DGPCMS

Step 1: **for** k =1 to M
Step 2:   Find a minimum spanning forest $F_k$ by interconnecting the end points of segments in R with (k-1) inter segments connectors. Let $T_1, T_2, \cdots, T_{(M-k+1)}$ be the tree components of $F_k$.
Step 3: $N_k$=0  /* Number of data mules used to collect data from k Euler paths */
Step 4:    **for** i =1 to (M-k+1)
Step 5:         $ST_i = \left\lceil \frac{2L(T_i)}{Vt} \right\rceil$     /* Number of sub paths for $k^{th}$ Euler path */
Step 6:         $N_k = N_k + 2*ST_i$
Step 7:   **end for**
Step 8: **end for**

Step 9: Let J be the index in between 1 to M such that $N_J$= min{$N_1, N_2, \cdots, N_M$}
Step 10: Construct Euler path $E_i$ for each tree $T_i \in F_J$.
Step 11: **for** i =1 to J
Step 12:    Partition the Euler path $E_i$ into $ST_i = \left\lceil \frac{L(E_i)}{Vt} \right\rceil$   segments of equal length and deploy two data mules at two ends of all segments.
Step 13: **end for**
Step 14: Two data mules from each segment will move inward direction in synchronous, when they meet each other's then reverse their movement direction to outward until they reach their starting positions again and continue the same process repeatedly.

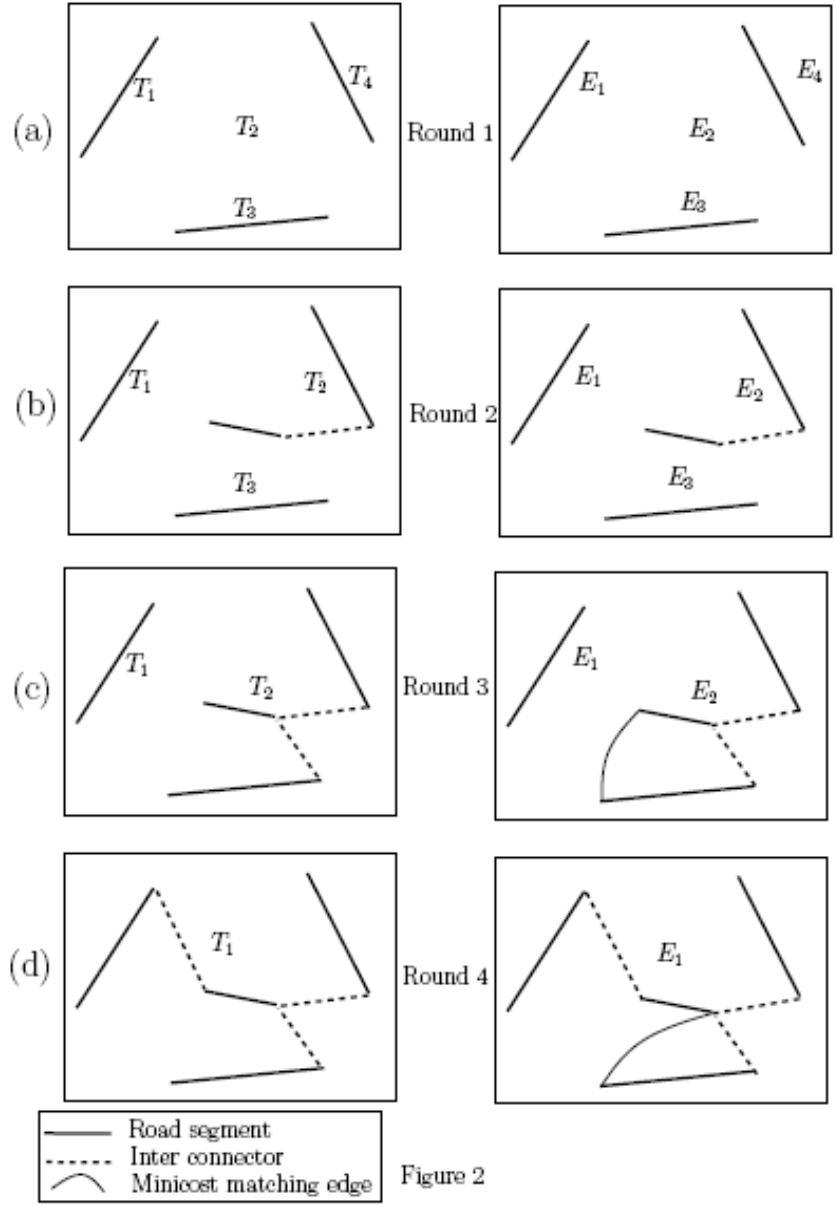

Figure 2

Sensor network in Figure 1 has four segments, four rounds for determining the usage of data mules is show in Figure 2. Initially in round 1, all the road segments are considered as independent tree as shown in Figure 2(a). Trees {$T_1$, $T_2$, $T_3$, $T_4$} are converted to Euler paths {$E_1$, $E_2$, $E_3$, $E_4$}. Determine the number of data mules requirement $N_1 = 2\sum_{i=1}^{4}\left\lceil\frac{L(E_i)}{Vt}\right\rceil \leq 2\sum_{i=1}^{4}\left\lceil\frac{2L(T_i)}{Vt}\right\rceil$, where $L(E_i)$ denotes length of Euler path $E_i$ and $Vt$ denotes the distance travel by data mule in time t. In round 2, reduce the number of trees by inter connecting closest pair of trees {$T_2$, $T_4$} of round 1 (connected by a dotted inter segment connector) and the new tree is referred as $T_2$. In this round, there are three trees {$T_1$, $T_2$, $T_3$} and their corresponding Euler paths $E_1$, $E_2$, $E_3$ are shown in Figure 2(b). Repeat the same process for round 3 and 4 with two trees and subsequently for one tree which are shown in Figure 2(c) and 2(d) respectively. Let J be the index of a round for which number of data mules requirement is minimum ($N_J$ = Min{$N_1$,$N_2$,$N_3$,$N_4$}). According to our algorithm, partition the Euler paths of the $J^{th}$ round into segments of equal length at most size $Vt$. An example of movements of data mules to collect data from mobile sensors from an Euler path is shown in Figure 3. In Figure 3(a), Euler path is split into three sub segments by two perpendicular segments. And for every segment two data mules are deployed at the two end points; and they start moving towards each other at the starting time t'=0. At time t'=t/2 the two data mules meet with each other at middle of the segment and change their movement to opposite directions as shown in Figure 3(b).

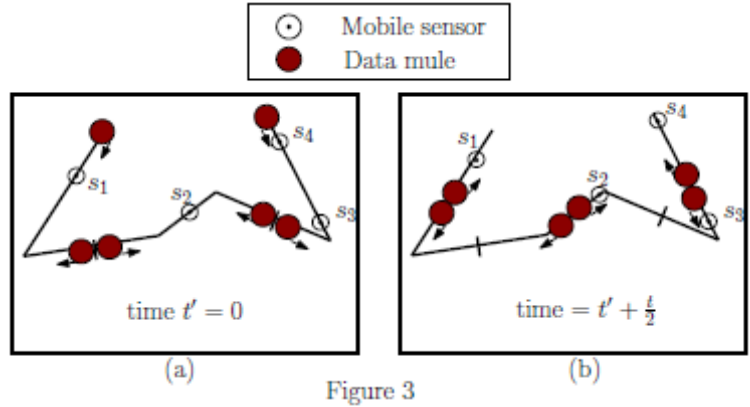

Figure 3

## 4.1. ANALYSIS

**Theorem 1**: Algorithm DGPCMS ensures that each mobile sensor is visited by a data mule at least once in every t time period.

**Proof:** Let at time t', a mobile sensor node belongs in one of the segment which is periodically traversed by its two corresponding data mules. Since the length of the segments is at most Vt and it is bounded by two data mules from two ends. The mobile sensor can't escape the segment without visited by its corresponding two data mules. All points of the segments are also visited by data mules with the interval t. Hence, the theorem is proved.

**Theorem 2**: Time complexity of the algorithm DGPCMS is $O(M^3)$.

**Proof:** DGPCMS algorithm determines a minimum length forest of k trees where k varying from 1 to M. It also determines the upper bound of the number of mules required to traverse Euler graphs corresponding to the k trees. Therefore, time complexity from Step 1 to Step 8 is $O(MlogM+M^2)$. Step 9 finding the minimum among M values can be determined in $O(M)$ time. Step 10 constructing Euler graph for all tree using minimum cost matching can be done on $O(M^3)$ time. Time complexity for Step 11 to Step 13 take $O(M)$ time. Hence the total time complexity of DGPCMS in $O(M^3)$.

**Theorem 3**: Number of data mules used in algorithm DGPCMS is $\leq 4OPT$. (OPT denotes the minimum number of data mules require to collect data from all mobile sensors.)

**Proof:** In the worst case, the data mules together must visit the full spanning forest within t time period. Since the speed of the data mule is V, therefore the numbers of data mules require to collect the data from the mobile sensors OPT$\geq \sum_{i=1}^{J} \left\lceil \frac{L(T_i)}{Vt} \right\rceil$, where $L(T_i)$ denotes the length of the tree $T_i$ and according to the algorithm DGPCMS, J is the index of the round for which the number of data mules used is minimum. Algorithm DGPCMS uses N=$2\sum_{i=1}^{J} \left\lceil \frac{L(E_i)}{Vt} \right\rceil \leq 2\sum_{i=1}^{J} \left\lceil \frac{2L(T_i)}{Vt} \right\rceil \leq 2\sum_{i=1}^{J} 2\left\lceil \frac{L(T_i)}{Vt} \right\rceil$ data mules to collect data from all mobile sensors. Therefore, N $\leq 4\sum_{i=1}^{J} \left\lceil \frac{L(T_i)}{Vt} \right\rceil \leq 4OPT$.

## 6. CONCLUSION

In this paper, we have proposed an approximation algorithm for data gathering protocol from mobile sensors using mobile data sinks. Our proposed algorithm overcomes the limitation in [9]. The algorithm will return a solution which within 4-factor of the optimal solution which will runs in $(M^3)$ time. In future, we will extend the work by relaxing the paths of the mobile sensors from line segments to an arbitrary region, where the mobile sensors can move arbitrarily within the bounded region.